\documentclass[twocolumn,showpacs,amsmath,amssymb,prl,superscriptaddress,floatfix]{revtex4}
\usepackage{graphicx}% Include figure files
\usepackage{bm}% bold math
\usepackage{amsmath,amsfonts}

\begin{document}

\title{Trap assisted creation of giant molecules and Rydberg-mediated coherent charge transfer in a Penning trap}

\author{I. Lesanovsky}
\affiliation{Institute for
Theoretical Physics, University of Innsbruck, and Institute for
Quantum Optics and Quantum Information of the Austrian Academy of
Sciences, Innsbruck, Austria}
\author{M. M\"{u}ller}
\affiliation{Institute for
Theoretical Physics, University of Innsbruck, and Institute for
Quantum Optics and Quantum Information of the Austrian Academy of
Sciences, Innsbruck, Austria}
\author{P. Zoller}
\affiliation{Institute for
Theoretical Physics, University of Innsbruck, and Institute for
Quantum Optics and Quantum Information of the Austrian Academy of
Sciences, Innsbruck, Austria}

\date{\today}

\begin{abstract}\label{txt:abstract}
We study two ions confined in a Penning trap. We show that electronically highly excited states exist in which an electron is delocalized among the two ions forming a giant molecule of several micrometer size. At energies close to the top of the Coulomb barrier these molecular states can be regarded as superpositions of Rydberg states of individual ions. We illuminate the possibility to observe coherent charge transfer between the ions. Beyond a critical principal quantum number the electron can coherently tunnel through the Coulomb barrier to an adjacent doubly charged ion. The tunneling occurs on timescales on which the dynamics of the nuclei can be considered frozen and radiative decay can be neglected.
\end{abstract}
\maketitle

As the simplest conceivable molecule, singly ionized molecular hydrogen $\mathrm{H}^+_2$ is formed out of two protons and a single electron. The electron is delocalized among the protons tying them together with an equilibrium distance of just a few Bohr's radii ($a_0$). In 2000, Greene \textit{et al.} complemented this class of conventional homonuclear binary molecules by predicting a new type of ultra-long-range molecules \cite{Greene00} in which a Rydberg atom forms a molecular entity with a ground state atom. Unlike in the former case the electron is delocalized over two nuclei which are separated by 1000 or more $a_0$. More recently, it has been shown that Rydberg atoms can form molecular states even if their electronic wave functions do not overlap \cite{Boisseau02,Farooqi03}. The atoms are held together by the multipole-interaction of their charge distributions giving rise to equilibrium distances of again several thousand $a_0$. Another class of 'molecular species' which however can only exist in a trap is constituted by trapped ions \cite{Leibfried03,Mitchell98}. At sufficiently low temperatures the mutual Coulomb repulsion which is compensated by the trapping potential leads to the formation of a Wigner crystal with the interparticle spacing being typically of the order of $\mu$m. The ions can be cooled to their motional ground state such that the low energy excitations are small oscillations around their equilibrium positions. These phonons are reminiscent of vibrational molecular excitations. However, unlike for the previously mentioned molecular species there are no delocalized electrons in the system. In this work we show theoretically the possibility to create delocalized electronic states among ions which are confined in a Penning trap \cite{Brown86,Mitchell98}. Such molecular states possess a spatial extension of several micrometers thereby constituting a novel class of trap induced quantum objects.

We focus on a setup consisting of two ions in a Penning trap which is constituted by an electric quadrupole and a homogeneous magnetic field. In a high magnetic field, which we consider here, the electronic motion is essentially one-dimensional and channeled along the magnetic field axis. In particular, we investigate states whose energy lies just below the Coulomb barrier separating the ionic charges. Such states can be thought of as being a superposition of states in which a single electron occupies a Rydberg state that can be associated with an individual ion. Along this line we study coherent charge transfer between two ions, i.e. the reaction $  A^{++} + A^+ \rightarrow A^+ + A^{++}$, which can be understood as a coherent tunneling process between the Coulomb wells. The transfer is initiated by transferring the electron of the singly charged ion to a high lying Rydberg state via a laser. We find tunneling rates of the order of hundred MHz.
Over the characteristic time the ionic motion can be considered frozen and radiative decay plays no role. We discuss the dynamics of the charge transfer as a function of the parameters of the Rydberg excitation laser and the tunneling rate of the electron.

The Penning trap is constituted by a homogeneous magnetic field $\mathbf{B}=B\,\mathbf{e}_z$ of strength $B$ and an electric quadrupole field with the potential $\Phi(\mathbf{r})= -\beta\left[x^2+y^2-2z^2\right]$ with $\beta$ being the field gradient. In the symmetric gauge, the vector potential is $\mathbf{A}(\mathbf{r})=\frac{1}{2} \mathbf{B} \times \mathbf{r}$ and the motion of a particle of charge $Qe$ and mass $M$ in this trap is governed by the Hamiltonian
\begin{eqnarray}
  H(\mathbf{p},\mathbf{r};Q,M)&=&\frac{1}{2M}\left[\mathbf{p}-Qe \mathbf{A}(\mathbf{r})\right]^2+Qe\,\Phi(\mathbf{r})\\
  &=&\frac{\mathbf{p}^2}{2 M}-\frac{\omega_c}{2} L_z +\frac{1}{2} M\left[\omega_\rho^2 \rho^2+\omega_z^2 z^2\right].\nonumber
\end{eqnarray}
Here $L_z=x p_y-y p_z=-i\hbar\partial_\phi$ is the $z$-component of the total angular momentum, and we have introduced the radial $\omega_\rho=\sqrt{(2/M)[(Q^2e^2B^2)/(8M)-Qe\beta]}$ and longitudinal $\omega_z=\sqrt{(Qe\beta)/M}$ trap frequencies as well as the cyclotron frequency $\omega_c=(Qe B)/M$.
In order to provide radial confinement $\omega_\rho$ has to be real, which is achieved beyond the critical field strength $B_1=\sqrt{(8M\beta)/(Qe)}$. We intend to operate the Penning trap in a regime where $\omega_\rho > \omega_z$, thus forming a prolate trap \cite{Crick08}. This configuration, in which the trapped ions line up in a chain is achieved for $B>B_2=4\sqrt{(M\beta)/(Qe)}$.

In the following we will focus on a scenario in which two ions that are of the same atomic species, but carry different charges, are placed in the trap. The feasibility of preparing and cooling two ion systems in a Penning trap has been experimentally demonstrated in Ref.~\cite{Crick08}. The Hamiltonian for the case in which \emph{the two ions are in their electronic ground state} reads $H=H(\mathbf{P}_1,\mathbf{R}_1;Q_1,M)+H(\mathbf{P}_2,\mathbf{R}_2;Q_2,M)
+V_\mathrm{C}(|\mathbf{R}_1-\mathbf{R}_2|;Q_1,Q_2)$
with the Coulomb interaction $V_\mathrm{C}(|\mathbf{R}_1-\mathbf{R}_2|;Q_1,Q_2)=(Q_1 Q_2e^2)/(4\pi\epsilon_0|\mathbf{R}_1-\mathbf{R}_2|)$. We assume that the ionic motion has been cooled to the ground state such that the ions perform harmonic oscillations around their equilibrium positions, which are given by
\begin{eqnarray}
  \mathbf{R}_{0,k}&=&\frac{\zeta}{2}\left(0,0,\frac{(1-k)Q_1+(2-k)Q_2}{(Q_1+Q_2)^{2/3}}\right),\label{eq:eq_positions}
\end{eqnarray}
with the characteristic length $\zeta=\left[e/(2\pi\epsilon_0\beta)\right]^{1/3}$ and $k=1,2$. The transverse oscillation frequencies are given by the respective values of $\omega_\rho$ whereas the frequencies of the longitudinal modes evaluate to $\omega^2_{z,\pm}=(2e\beta)/(M(Q_1+Q_2))\left[Q_1^2+6Q_1Q_2+Q_2^2\pm\sqrt{Q_1^4+14Q_1^2Q_2^2+Q_2^4}\right]$.
Throughout this work we will consider $^{\mathrm{40}}\mathrm{Ca}$ with $Q_1=2$ and $Q_2=1$ confined to a Penning trap with the parameters $B=10\,\mathrm{T}$ and $\beta=2\times 10^6\,\mathrm{V}/\mathrm{m}^2$. In this case we find $\omega_{\rho,1}=2\pi\times 1.85\,\mathrm{MHz}$, $\omega_{\rho,2}=2\pi\times 3.78\,\mathrm{MHz}$, $\omega_{z,-}=2\pi\times 0.83\, \mathrm{MHz}$ and $\omega_{z,+}=2\pi\times 1.44\, \mathrm{MHz}$. We furthermore have the characteristic length $\zeta=11.29\,\mu\mathrm{m}$ and consequently an interionic distance of $D=8.14\,\mu\mathrm{m}=153\times 10^3\,a_0$ with $a_0$ being Bohr's radius.

The idea of creating delocalized electronic states in this setup goes as follows: The valence electron of the singly charged ion is excited by a laser to a high lying electronic (Rydberg) state. Here the electron is exposed to the potential of \emph{two doubly charged ions}. The electron can hop as soon as its wave function overlaps with the empty Ryd\-berg orbital of the second ion. By this electronic states that extend over both ionic cores are formed and the electronic charge can be transferred. The size of a Rydberg orbit is approximately given by $r_\mathrm{Ryd}=(2a_0\,n^2)/Q $ with $n$ being the principal quantum number. Requiring $r_\mathrm{Ryd}$ to be half the equilibrium distance $D$ between the ions one finds that delocalized electronic states are expected to occur beyond $n>n_\mathrm{ct}=\sqrt{D/(2a_0)}\approx 277$ for the above parameters. Indeed, we will see later that this is the case even for much smaller values of $n$.

In order to study the process more quantitatively, we consider the electronic Hamiltonian \begin{eqnarray}
  H_\mathrm{el}&=&H(\mathbf{p},\mathbf{r};-1,m)\label{eq:el_hamiltonian_3d}\\
  &&+V_\mathrm{C}(|\mathbf{r}-\mathbf{R}_{0,1}|;2,-1)+V_\mathrm{C}(|\mathbf{r}-\mathbf{R}_{0,2}|;2,-1).\nonumber
\end{eqnarray}
The ions are frozen at their equilibrium position as we are interested in the electronic dynamics on a shorter timescale than the ionic motion. Moreover, the electronic spin is not considered as it merely gives rise to a constant energy offset. In Eq.~(\ref{eq:el_hamiltonian_3d}) we have approximated the interaction potential between the ionic cores and the electron by a sum of two pure Coulomb potentials of charge $2e$. This is a simplification, since in general the true potential will deviate from the Coulomb potential at small distances due to the presence of the inner electronic shells. For Rydberg states this gives rise to the quantum defect of states with low angular momenta. The general features of the charge transfer, however, will not be affected by the actual value of the quantum defect and an experimental implementation requires in any case a careful spectroscopic analysis.

In the following, we will motivate an adiabatic single channel approximation in order to
derive an effective one dimensional electronic Hamiltonian. To this end it is instructive to consider the part of $H(\mathbf{p},\mathbf{r};-1,m)$ in Eq.~(\ref{eq:el_hamiltonian_3d}) which governs the transversal electronic dynamics, providing harmonic confinement with the frequency $\Omega_\rho=\sqrt{(2/m)\left[(e^2B^2)/(8m)+e\beta\right]}=2\pi\times 140\,\mathrm{GHz}$. The eigenstates of this potential are given by the two-dimensional harmonic oscillator states $\Theta_{\nu,\mu}(\rho,\phi)$ with the energies $E_{\nu,\mu}= 2\hbar\Omega_c \mu +\hbar \Omega_{\rho}\left(2\nu+|\mu|+1\right)$, where $\Omega_c=(e B)/m$. We will see a posteriori that in the regime of interest the energy gap between the states $\Theta_{\nu,\mu}(\rho,\phi)$ and $\Theta_{\nu\pm 1,\mu}(\rho,\phi)$ is much larger than the energy of the longitudinal electronic motion. Therefore we can neglect coupling between oscillator states with different $\nu$. For fixed $\nu$, states with different $\mu$ are only coupled if the interatomic axis does not coincide with the $z$-axis. For infinitely heavy ions in equilibrium this is not the case. In the case of finite mass, the corresponding couplings will lead to a modification of the electronic wave function depending on the positions of the ions which has to be taken into account by averaging over the ionic probability density. The dominant part of this averaging however will arise from the configuration where both ions are located on the $z$-axis. We therefore can neglect couplings between states with different $\mu$ and still obtain a qualitatively correct picture. In our approximation the transversal electronic motion is frozen in the oscillator ground state $\Theta_{0,0}(\rho,\phi)=(\sqrt{\pi}\rho_0)^{-1}\exp\left[-(\rho/\rho_0)^2/2\right]$ with $\rho_0=\sqrt{\hbar/(m\Omega_\rho)}$ and we use the ansatz $\Psi(\rho,z,\phi)=\Theta_{0,0}(\rho,\phi)\phi(z)$ for the electronic eigenstates.
Such quasi one-dimensional scenario, in which the electronic motion is channeled along the magnetic field axis is typical for atoms and molecules \cite{Turbiner06} in strong magnetic fields. It has also been reported for electron-hole pairs in a semiconductor at high magnetic fields \cite{Elliot60} and for long-range molecules formed by a ground state and a Rydberg atom \cite{Lesanovsky06}. The effective Hamiltonian for the longitudinal dynamics is obtained by $H_\parallel=\int d\phi\int d\rho\,\rho\, \Theta_{0,0}^*(\rho,\phi)H_\mathrm{el}\Theta_{0,0}(\rho,\phi)$. Neglecting a constant energy offset one obtains
\begin{eqnarray}
  H_\parallel&=&\frac{p_z^2}{2m}-\frac{e^2}{2\pi\epsilon_0}\frac{\sqrt{\pi}}{\rho_0}\sum_{k=1}^2 \exp\left[(z-Z_{0,k})^2/\rho_0^2\right]\nonumber\\
  &&\qquad\qquad\times\mathrm{erfc}\left(\frac{|z-Z_{0,k}|}{\rho_0}\right)-\frac{1}{2}m\Omega^2_z z^2\label{eq:H_1_d}
\end{eqnarray}
where $\Omega_z=2\sqrt{(e\beta)/m}=2\pi\times 189\,\mathrm{MHz}$ is the 'frequency' of the inverted parabolic potential and $\mathrm{erfc}(x)$ is the complementary error function.
\begin{figure}\center
\includegraphics[scale=0.4]{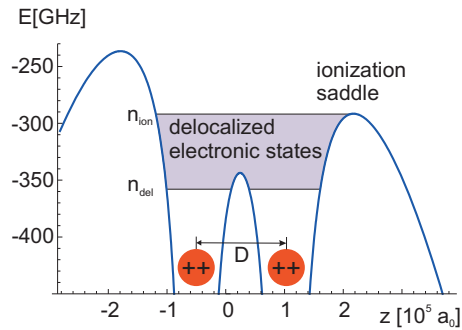}
\caption{Electronic potential if the ions are frozen at their equilibrium positions. The electronic potential supports bound states which are delocalized between the two nuclei. For too large excitation energies the saddle point due to the inverted harmonic potential leads to field ionization.} \label{fig:potential}
\end{figure}
The corresponding longitudinal potential is presented in Fig.~\ref{fig:potential}. If the electron is far away from either ionic core the potential is approximately a sum of two one-dimensional Coulomb potentials $V_\mathrm{1d}=-e^2/(2\pi\epsilon_0|z|)$ centered at the respective ionic equilibrium positions. For small excitation energies the potential wells are well separated. If the excitation energy surpasses $E_\mathrm{del}\approx - (2e^2)/(\pi\epsilon_0 D)$ the wells merge and hence electronic states which extend over both ionic cores are possible. Expressing $E_\mathrm{del}$ in terms of the principal quantum number $n$ one finds this merging to occur approximately at $n_\mathrm{del}=(e/\hbar)\sqrt{(m D)/(16\pi\epsilon_0)}$. For the employed parameter set one obtains $n_\mathrm{del}\approx 196.14$ which is significantly smaller than the principal quantum number we obtained earlier by invoking the argument of overlapping wave functions. Further inspection of the potential reveals that the inverted parabolic potential due to the Penning trap eventually leads to field ionization which is shown in Fig.~\ref{fig:potential}. The position of the saddle ($z_\mathrm{sd}>Z_{0,2}$) is given by $z_\mathrm{sd}=\gamma Z_{0,2}$ where $\gamma>1$ is the solution of the equation $(\gamma-1)^{-2}+(\gamma+1/2)^{-2}-\sigma \gamma=0$ and $\sigma=\left[(1/2)m \Omega_z^2 Z^2_{0,2}\right]\left[e^2/(4\pi \epsilon_0 |Z_{0,2}|)\right]^{-1}=(1/2)Q_1^3/(Q_1+Q_2)^2=4/9$. This equation is solved by $\gamma=2.12$ and thus the saddle point is for our parameters located at $z_\mathrm{sd}=217523\,a_0$. From this we find that classically ionization is expected to occur for principal quantum numbers beyond $n_\mathrm{ion}=1.456\sqrt{Z_{0,2}/a_0}=(0.21\,e^{2/3}\sqrt{m})/(\hbar \epsilon_0^{2/3}\beta^{1/6}) \approx 466$. Hence there is an energy window $n_\mathrm{del}<n<n_\mathrm{ion}$ in which delocalized electronic states exist. Here the escape of the electron from the trap is prevented by its attraction to the doubly charged ion. For the given parameters the width of this window is about $60\,\mathrm{GHz}$.

We now focus on the electronic states which lie close to the top of the Coulomb barrier. In this regime the influence of the anti-confining potential can be approximately neglected and the problem assumes a particularly symmetric appearance. We obtain the eigenstates of interest by numerically integrating the Schr\"odinger equation of the Hamiltonian (\ref{eq:H_1_d}) with $\Omega_z=0$ in the vicinity of $n_\mathrm{del}$. With respect to reflections at the symmetry point $z_\mathrm{sym}=(Z_{0,1}+Z_{0,2})/2$ we can define gerade and ungerade states.
\begin{figure}\center
\includegraphics[scale=0.55]{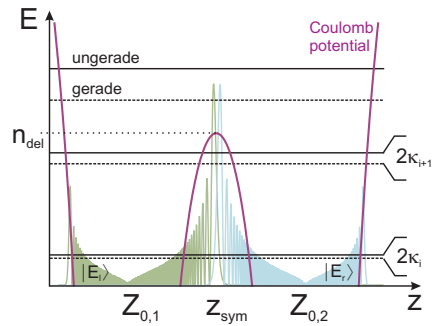}
\caption{Sketch of the energy level structure close to $n_\mathrm{del}$. Here the inverted parabolic potential can be neglected and the states can be characterized by their symmetry property under reflection at $z_\mathrm{sym}$. The energies of these pairs of gerade/ungerade states are indicated by dashed/solid lines. The pairs are almost degenerate with a small energy splitting $\kappa_i$. Wave functions which are localized in one of the two wells ($\left|E_{l/r}\right>$) can be created by linear superpositions of the gerade and ungerade states. Two such wave functions are exemplarily shown.} \label{fig:tunneling}
\end{figure}
In Fig.~\ref{fig:tunneling} we show a sketch of the energy level structure in the vicinity of $n_\mathrm{del}$. Gerade ($\left|E_g\right>$) and ungerade states ($\left|E_u\right>$) are energetically split by the energy $2\kappa=E_u-E_g$ with $\kappa$ being the tunnel coupling. The splitting increases with increasing degree of excitation. States which are localized in the left ($\left|E_l\right>$) or right well ($\left|E_r\right>$) can be created by the linear combinations $\left|E_l\right>=(1/\sqrt{2})\left[\left|E_g\right>+\left|E_u\right>\right]$ and $\left|E_r\right>=(1/\sqrt{2})\left[\left|E_g\right>-\left|E_u\right>\right]$, respectively. They are not stationary but evolve under the Hamiltonian $H_\mathrm{tunnel}=\kappa\left|E_l\right>\left<E_r\right|+\mathrm{h.c.}$ which describes Rabi oscillations of the electronic density between the left and right well at a rate $\kappa$. As an example we provide two sets of gerade/ungerade eigenstates which were obtained for the above parameters: Their energies correspond to the principal quantum numbers $n^1_g=195.6848$, $n^1_u=195.6917$ and $n^2_g=196.0161$, $n^2_u=196.0662$ ($E(n)=-4\times 13.6\,\mathrm{eV}n^{-2}$). The fact that these values are smaller than $n_\mathrm{del}$ indicates that the electron in the respective states indeed tunnels through the barrier which separates the two Coulomb wells. The tunnel coupling in the two cases evaluates to $\kappa_1/\hbar=2\pi\times 12.3 \,\mathrm{MHz}$ and $\kappa_2/\hbar=2\pi\times 87.5\,\mathrm{MHz}$. Since these rates are significantly larger than the oscillation frequencies of the ionic cores it is indeed justified to use the picture of a frozen ionic motion. Moreover, the rates are much larger than the radiative decay rate of Rydberg states which is usually smaller than $1\,\mathrm{MHz}$. Therefore the described process can be considered coherent.

\begin{figure}\center
\includegraphics[scale=0.35]{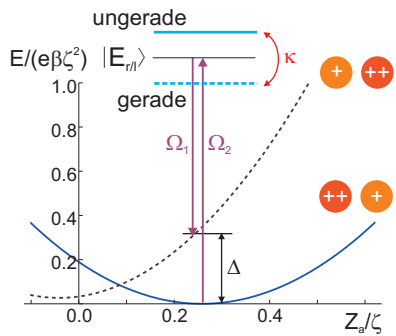}
\caption{Cut through the potential curves along the normal coordinate $Z_a$ with $Z_b=0$. The equilibrium positions of the initial configuration ($A^{++}+A^+$) do not coincide with those of the final configuration ($A^{+}+A^{++}$). As a consequence there is a relative detuning $\Delta$ between the initial and final state which has to be accounted for during the laser excitation.} \label{fig:transfer}
\end{figure}

We are now in position to study the charge transfer process $A^{++}+A^+\rightarrow A^{+}+A^{++}$. To this end we employ two laser fields coupling the ground state of the right ion $\left|G_r\right>$ resonantly to the state $\left|E_r\right>$ with a Rabi frequency $\Omega_2$. Due to the tunnel coupling the state $\left|E_r\right>$ evolves into the state $\left|E_l\right>$ which is then resonantly coupled to the ground state $\left|G_l\right>$ of the left ion with Rabi frequency $\Omega_1$. While initially the two ions resided in their equilibrium position this is no longer true for the final state as here the charges are interchanged. The new equilibrium positions can be calculated by Eqs.~(\ref{eq:eq_positions}). However, if the ions are frozen, their energy in the final state, i.e. $A^{+}+A^{++}$, is higher than in equilibrium. Hence, one has to account for this energy off-set $\Delta$ in order to make the transition $\left|E_l\right>\rightarrow \left|G_l\right>$ resonant. This is shown in Fig.~\ref{fig:transfer} where the potentials of the ground state configurations $A^{++}+A^+$ and $A^{+}+A^{++}$ are depicted. The intersections shown are made along the normal coordinate $Z_a=\alpha_- Z_1+\alpha_+ Z_2$ at $Z_b=-\alpha_+ Z_1+\alpha_- Z_2$ set to zero. The coefficients are given by $\alpha_\pm=\sqrt{(1/2)\pm3/(2\sqrt{73})}$. The energy gap between the two configurations evaluates to $\Delta=(e\beta\zeta^2)/(2\times 3^{1/3})$ which yields for our parameters $\Delta=2\pi\times 844\,\mathrm{GHz}$.

\begin{figure}\center
\includegraphics[scale=0.32]{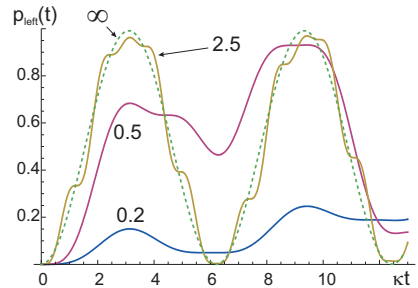}
\caption{Time dependence of the probability to find the electron with the left ion for $\delta=0.2,\,0.5,\, 2.5$ and the asymptotic curve for $\delta\gg 1$. Initially the electron is with the right ion. The time-evolution is governed by the Hamiltonian (\ref{eq:model_hamiltonian}).} \label{fig:probability}
\end{figure}
With the two lasers turned on and choosing $\Omega_1=\Omega_2=\Omega$ the electronic dynamics is governed by the Hamiltonian
\begin{eqnarray}
  H_\mathrm{tunnel}=\Omega \left|G_l\right>\left<E_l\right|+\Omega \left|G_r\right>\left<E_r\right|+\kappa\left|E_l\right>\left<E_r\right|+\mathrm{h.c.},\label{eq:model_hamiltonian}
\end{eqnarray}
where $\left|G_{r/l}\right>$ is the ground state of a \emph{singly charged} ion at the left or right position in the trap. The probability of finding the electron in the left potential well is given by
\begin{eqnarray}
  p_\mathrm{left}(t)=\frac{1}{1+4\delta^2}\left[\delta^2(\cos(\epsilon_- \kappa t)-\cos(\epsilon_+ \kappa t))^2\right.\nonumber\\
  \left.+(\epsilon_+\sin(\epsilon_- \kappa t)-\epsilon_-\sin(\epsilon_+ \kappa t))^2\right]
\end{eqnarray}
with $\epsilon_\pm=\sqrt{(1+2\delta\pm\sqrt{1+4\delta^2})/2}$ and $\delta=\Omega/\kappa$.
In Fig.~\ref{fig:probability} this probability is shown for three different values of $\delta$. For $\delta=0.2$ the electron mainly stays with the right ion over the time interval shown. As $\delta$ increases the likelihood to find the electron in the left well increases significantly for small times. In the limit $\delta\gg 1$ one finds $p_\mathrm{left}(t)=(1/2)(1-\cos(\kappa t))$ and the electron is found with certainty to be with the left ion at the times $t_m=(2m+1) \pi/\kappa$. Once the transfer is performed and the lasers are switched off the system is in a highly excited state as can be seen from the potential surfaces presented in Fig.~\ref{fig:transfer}. A successful charge transfer will be accompanied by strong oscillations of the ions. After cooling the ionic motion the interchange of the charge can be monitored by fluorescence imaging on a frequency which is only resonant on a transition of the singly charged ion. The above treatment is only valid for time intervals over which the ions don't move. Moving ions would turn the tunneling rate into a dynamic quantity.

In principle the presented scenario can be extended to more than two ions. An electron could then form states which are delocalized over an entire ion chain. Such system would constitute an implementation of the tight-binding model. Future directions of this work are to create and to study one- and two-dimensional realizations of such system and to include more than one electron in the dynamics which would give rise to Hubbard-models.

We gratefully acknowledge discussions with H. Sadeghpour, P. O. Schmidt and H. H\"affner. I.L. acknowledges the hospitality of ITAMP where fractions of this work were completed. We acknowledge support from the Austrian Science Fund (FWF) within the Spezialforschungsbereich 15,
from the European Union within the SCALA network and from the Institute for Quantum Information.

\end{document}